\documentclass[conference]{IEEEtran}
\usepackage{booktabs}			
\usepackage{tabulary}			
\usepackage{cite}
\usepackage{url}
\usepackage{breakurl}
\usepackage[breaklinks]{hyperref}

\ifCLASSINFOpdf
  \usepackage[pdftex]{graphicx}
  \DeclareGraphicsExtensions{.pdf,.jpeg,.png,.eps}
\else
\fi
\begin{document}

\title{Enron versus EUSES: \\
A Comparison of Two Spreadsheet Corpora}
\author{\IEEEauthorblockN{Bas Jansen}
\IEEEauthorblockA{Delft University of Technology\\Email: b.jansen@tudelft.nl}}
\maketitle

\begin{abstract}
Spreadsheets are widely used within companies and often form the basis for business decisions. Numerous cases are known where incorrect information in spreadsheets lead to incorrect decisions. Such cases underline the relevance of research on the professional use of spreadsheets.

Recently a new dataset became available for research, containing over 15.000 business spreadsheets that were extracted from the Enron E-mail Archive. With this dataset, we 1) aim to obtain a thorough understanding of the characteristics of spreadsheets used within companies, and 2) compare the characteristics of the Enron spreadsheets with the EUSES corpus which is the existing state of the art set of spreadsheets that is frequently used in spreadsheet studies.

Our analysis shows that 1) the majority of spreadsheets are not large in terms of worksheets and formulas, do not have a high degree of coupling, and their formulas are relatively simple; 2) the spreadsheets from the EUSES corpus are, with respect to the measured characteristics, quite similar to the Enron spreadsheets.
\end{abstract}

\section{Introduction}

Spreadsheets are used widely within companies. It was, for example, estimated that 95\% of U.S. firms use spreadsheets for financial reporting \cite{Panko:2008aa}. The information in spreadsheets often forms the basis for significant business decisions\cite{hermans2011supporting}. However, from previous research we know that spreadsheets are error-prone \cite{panko1998we}. This poses the risk for companies of taking important decisions on inaccurate information. Eventually, this can lead to incorrect decisions and loss of money. The growing list of Horror Stories of the European Spreadsheet Risk Interest Group\footnote{www.eusprigorg/horror-stories.htm} illustrates this and stresses the importance of spreadsheet research. It is important to understand why spreadsheet errors are easily made and how this can be improved.

One way to address this, is to study a large set of real world spreadsheets. The EUSES corpus, a set of approximately 4,500 spreadsheets from different domains, is used most frequently for this purpose. Although the EUSES corpus has proven to be very useful and is used in many research projects \cite{abraham2007goaldebug, hermans2012detecting, dou2014spreadsheet}, there are also some drawbacks. First of all the majority of these spreadsheets were obtained from the public world-wide-web. We do not know if these spreadsheets are similar to spreadsheets that are used within companies. Furthermore the size of the corpus is relatively small. Finally we do not know anything about the context in which the spreadsheets were created.

Previously, attempts have been made to compose a dataset of spreadsheets that are used within companies, but it is very difficult to convince companies to share their spreadsheets for research purposes. And if they are willing to do so, it is even more difficult to get permission to make these spreadsheets available for the research community.

Recently Hermans and Murphy-Hill introduced a new set of spreadsheets \cite{Hermans:2014aa}, consisting of more than 15,000 spreadsheets that were extracted from the already existing Enron Email archive that was made public during the legal investigation following the Enron bankruptcy \cite{klimt2004introducing}. The size of this dataset is significantly larger than the EUSES corpus and maybe even more important: we know that these spreadsheets were used in industry. This dataset is publicly available\footnote{www.felienne.com/enron}.

Hermans and Murphy-Hill performed a preliminary analysis of some basic characteristics (like number of worksheets, number of cells, etc.) of these spreadsheets. To obtain a more thorough understanding of the characteristics of industry grade spreadsheets, we extend the analysis with additional metrics on the degree of coupling and the use of functions. Furthermore, we compare this dataset extensively with the EUSES corpus and, by using statistical tests, answer the question of its representativeness with respect to spreadsheets that are used by companies.

The contributions of this paper are:
\begin{itemize}
\item A detailed analysis of a large dataset of industrial spreadsheets on dimensions of size and coupling
\item A detailed analysis of how functions are used in spreadsheets in a business environment
\item A comparison of the characteristics of the spreadsheets in the Enron corpus with the EUSES corpus.
\end{itemize}

\section{Research Questions}

The Enron dataset allows us to understand and quantify the characteristics of industry grade spreadsheets. In this study we focus on the level of coupling and the size of spreadsheets. The reason for this is that we know from software engineering that source code tends to be more error-prone when the coupling between the different units of a program is high or the program consists of large parts \cite{fowler1999refactoring}. Previous research indicates that this also applies to spreadsheets \cite{hermans2014detecting}.

Modern spreadsheet systems contain over 300 different functions and with these functions users can create very complex formulas. In previous work, we introduced the concept of a visual language for spreadsheets \cite{jansen2014using} for which functions form the building blocks. In this study we therefore analyze, in addition to the metrics on size and coupling, how functions are used.

To provide context, we compare the characteristics of spreadsheets that are used in industry with the spreadsheets of the EUSES corpus. It will also allow us to evaluate the representativeness of the EUSES corpus.

Hence we aim to answer the following research questions in this paper:
\begin{itemize}
\item What are the characteristics, in terms of size and coupling, of spreadsheets in a business environment?
\item How are spreadsheet functions used?
\item Is there a difference with respect to these metrics between spreadsheets that are used in a business context and the EUSES corpus?
\end{itemize}

\section{Approach}

For this paper, we use two datasets: Enron and EUSES. We analyzed the spreadsheets in the two datasets with the Spreadsheet Scantool, developed at Delft University of Technology. The tool runs on the previously developed Breviz core, that was made for spreadsheet visualization and smell detection \cite{hermans2012analyzing}. The Scantool collects several metrics on spreadsheet, worksheet and cell level. The metrics that were used for this paper can be found in Table \ref{overviewofusedmetrics}.

\begin{table}[htbp]
\begin{minipage}{\linewidth}
\setlength{\tymax}{\linewidth}
\centering
\small
\caption{Overview of used metrics}
\label{overviewofusedmetrics}
\begin{tabulary}{\textwidth}{lJ}\toprule
Dimension&Metric\\
\midrule
Size&s1 \# non-empty cells per spreadsheet\\
&s2 \# worksheets per spreadsheet\\
&s3 \# formulas per spreadsheet\\
&\textbf{s4 \# unique formulas per spreadsheet}\\
&\textbf{s5 length of formula in characters}\\
\midrule
Coupling&c1 \% spreadsheets linked to other spreadsheets\\
&c2 \# external links per spreadsheet\\
&c3 \% spreadsheets with intraworksheets connections\\
&c4 \# intraworksheet connections per spreadsheet\\
&c5 \# passing cells as a percentage of total cells\\
&\textbf{c6 path depth per formula}\\
&\textbf{c7 transitive precedents per formula}\\
\midrule
Functions&f1 usage of built-in functions\\
&f2 usage of functions per category\\
&\textbf{f3 \# precedents per formula}\\
&\textbf{f4 parse tree depth}\\
&f5 \# unique functions per formula\\
\bottomrule
\end{tabulary}
\end{minipage}
\end{table}

Some of these metrics (marked bold in Table \ref{overviewofusedmetrics}) deserve some further explanation.

\begin{itemize}
\item \textbf{Number of unique formulas per spreadsheet (s4)} In spreadsheets, it is very common to define a formula in one cell and then copy it down or right to other cells. For this reason, many of the formula cells in a spreadsheet contain the same formula and only the references to other cells differ. Therefore, we also measure the number of unique formulas in the spreadsheet. We do this on the level of the worksheet. Two identical formulas on different worksheets will count as two unique formulas. We determine the unique formulas by looking at the relative R1C1 notation of the formula. As indicated by Sajaniemi, this notation stays the same even if you copy it down or right \cite{sajaniemi2000modeling}.
\item \textbf{Length of formula (s5)} From software engineering, we know that a large number of lines of code increase the chance on errors. In spreadsheets, we could consider the individual formulas to be comparable to lines of code. So, if more lines of source code lead to a higher error rate, it is reasonable to assume that the longer the formula, the more likely it is that it contains an error. Therefore, we measure the length of formulas in characters.

\begin{figure}
\centering
\includegraphics[width=0.85\linewidth]{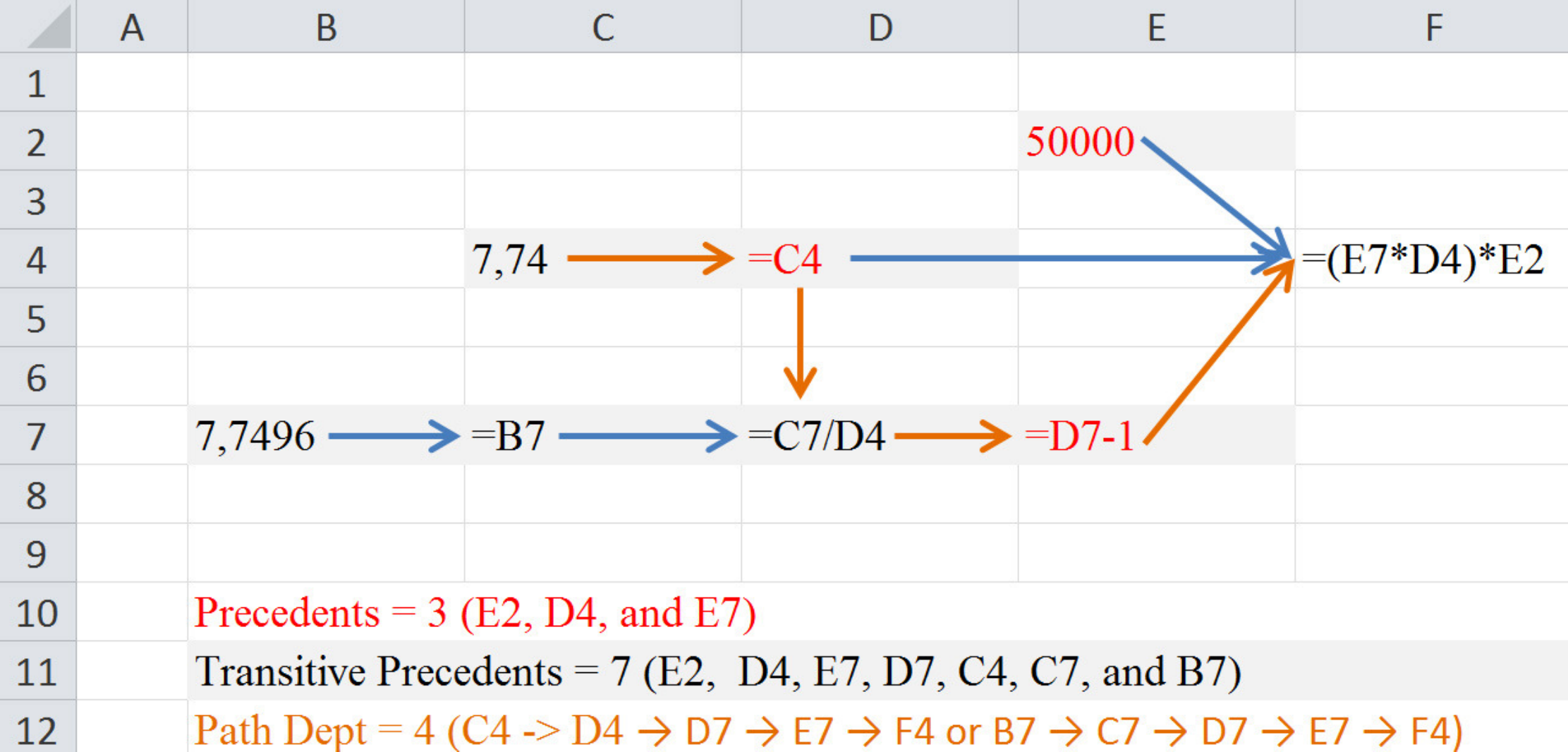}
\caption{Precedents, Transitive Precedents, and Path Depth}
\label{fig:Precedents}
\end{figure}

\item \textbf{Path depth (c6), transitive precedents (c7), and precedents (f3)} Most formulas receive input from other cells. These input cells are the so called precedents. These precedents themselves could also receive input from other cells. If you trace along these precedents until you reach a cell without one, you have the number of transitive precedents. The path depth is the longest calculation chain, see also Figure \ref{fig:Precedents}.
\item \textbf{Parse tree depth (f4)} This is a measure of how nested a formula is. The formula D9 + E9 has a parse tree dept of 2, the formula (B5 - T5) / (B6 * SQRT(4)) a parse tree depth of 5.
\end{itemize}

To determine if there is any difference in terms of these metrics between the Enron and EUSES spreadsheets, we have calculated the difference between the distributions of the two datasets using a Wilcoxon-Mann-Whitney test and the Cliff's Delta \emph{d} effect size. We use this test to analyze whether there is a significant difference between the the distribution of the metrics of the Enron and EUSES spreadsheets. If a significant difference is found, we use the Cliff's Delta effect size to measure the magnitude of the difference.

\section{Results}

In this section we present the results of our analysis of the Enron and EUSES spreadsheets. First we will discuss the metrics for size and coupling. Next we have a closer look at the use of functions and we conclude this section with a comparison between the Enron and the EUSES spreadsheets.

\subsection{Size}
Previous research has shown that lengthy source code will increase the error-rate\cite{fowler1999refactoring}. Therefore we analyse both datasets on the dimension of size. The results of this analysis can be found in Table \ref{sizemetrics}. Two measures for the size of a spreadsheet are the number of non-empty cells (s1) and the number of worksheets (s2)\footnote{Up to Excel 2010 every spreadsheet started by default with 3 workbooks. Because of this there are a lot of spreadsheets with 3 workbooks of which 2 are empty. For this reason we excluded empty worksheets to calculate s2} We notice that on average the Enron spreadsheets contain more non-empty cells than the EUSES spreadsheets. The number of worksheets per spreadsheet is almost identical for both datasets. For context consider a study of  Hermans \emph{et. al.}\cite{hermans2014detecting}. They asked financial professionals to select one of their large and complex spreadsheets. These spreadsheets contained tens of thousands of non-empty cells and have on average 10 worksheets. Based on this we can conclude that both number of cells and the number of worksheets in the Enron and EUSES spreadsheets are relatively small.

\begin{table*}[htbp]
\begin{minipage}{\linewidth}
\setlength{\tymax}{\linewidth}
\centering
\small
\caption{Overview of size metrics}
\label{sizemetrics}
\begin{tabulary}{\textwidth}{llrrrrr}\toprule
Metric&Dataset&Min&Q1&Median&Q3&Max\\
\midrule
s1 Number of non-empty cells per spreadsheet&Enron&7&204.0&701.0&3,672.0&889,952.0\\
&EUSES&14&237.0&573.5&1,354.3&113,134.0\\
\midrule
s2 Number of worksheets per spreadsheet&Enron&1.0&1.0&1.0&4.0&175.0\\
&EUSES&1.0&1.0&1.0&3.0&106.0\\
\midrule
s3 Number of formulas per spreadsheet&Enron&1.0&16.0&128.0&758.0&175,568.0\\
&EUSES&1.0&19.0&73.0&253.3&26,434.0\\
\midrule
s4 Number of unique formulas per spreadsheet&Enron&1.0&4.0&14.0&50.0&6,862.0\\
&EUSES&1.0&3.0&10.0&32.0&961,0\\
\midrule
s5 Length of formula in characters&Enron&1.0&12.0&18.0&29.0&1,111.0\\
&EUSES&1.0&14.0&20.0&35.0&1,129.0\\
\bottomrule
\end{tabulary}
\end{minipage}
\end{table*}

Formulas process and manipulate the data in the spreadsheet. So another way to look at the size of a spreadsheet is to look at 1) the number of formulas per spreadsheet (s3), 2) the number of unique formulas per spreadsheet (s4), and 3) the length of the formulas in characters (s5). Within the Enron set, formulas are used more frequently than in the EUSES set, but the latter tend to be slightly longer.

\subsection{Coupling}

Besides size, a high degree of coupling between the units of a program could also lead to more errors. We therefore analyze the degree of coupling in spreadsheets. This can be done on several levels within a spreadsheet:

\begin{itemize}
\item Coupling between spreadsheets
\item Coupling between worksheets
\item Coupling between cells
\end {itemize}

The values of the Enron and EUSES spreadsheets for the different measures of coupling are summarized in Table \ref{couplingmetrics}.

\subsubsection{Spreadsheets}
Starting with coupling between spreadsheets (c1), we can observe that only a minority of spreadsheets links to another spreadsheet (Enron 11\%, EUSES 1\%). For both sets, the majority of these spreadsheets only link to one or two other spreadsheets. The use of external links (c2) is slightly higher in the Enron set than the EUSES set.

\subsubsection{Worksheets}
On the worksheet level we observe that, although there are more spreadsheets with one or more intraworksheet connections (c3) than spreadsheets with external links, it is still a minority (Enron 20\%, EUSES 10\%). If we look at the number of connections (c4), we find that the median for Enron and EUSES is respectively four and three, see Table \ref{couplingmetrics}.

It is fair to state that the level of coupling on both the level of spreadsheets and the level of worksheets is low. The vast majority of spreadsheets contain no links to other spreadsheets or other worksheets within the same spreadsheet.
 
\subsubsection{Cells}
On the cell level, we have analyzed how formulas are linked to another via cell references. A formula receives values from other cells (precedents / cell fan in), and the result of the formula can be used in other formulas (dependents / cell fan out). To gain a better understanding of the degree of coupling on the cell level, we look at the path depth (longest path of precedents) and the total number of transitive precedents, see Table \ref{couplingmetrics}.

Before we can calculate the number of transitive precedents, there is one adjustment that we have to make. A pattern that we witness frequently in spreadsheets is a formula that is just passing data (eg. `=A1'). Within the Enron set this is true for 35\% of the formulas and for EUSES this is 22\% (c5). Most of these passing cells have a path depth and number of precedents of one.\footnote{The number of precedents could be higher than one if the formula is passing data from merged cells.} To get a better insight of the `real' links we have excluded these passing formulas in our analysis of path depth and number of precedents.

The path depth (c6) of Enron is almost identical to EUSES. Both data sets have a median path depth of one, which is small. If we look at the total number of cells that influence a formula (c7), we observe that the median for Enron and EUSES are relatively low and almost the same (respectively three and four). Noteworthy is the maximum of more than 22,000 transitive precedents for a single formula in the Enron set.

\begin{table*}[htbp]
\begin{minipage}{\linewidth}
\setlength{\tymax}{\linewidth}
\centering
\small
\caption{Overview of coupling metrics}
\label{couplingmetrics}
\begin{tabulary}{\textwidth}{llrrrrr}\toprule
Metric&Dataset&Min&Q1&Median&Q3&Max\\
\midrule
c2 Number of external links per spreadsheet&Enron&1.0&1.0&2.0&5.0&155.0\\
\scriptsize for spreadsheets with at least one external link&EUSES&1.0&1.0&1.0&2.0&88.0\\
\midrule
c4 Number of intraworksheet connections per spreadsheet&Enron&1.0&1.0&4.0&13.0&735.0\\
\scriptsize for spreadsheets with at least one intraworksheet connection&EUSES&1.0&1.0&3.0&7.0&121.0\\
\midrule
c6 Path depth per formula&Enron&0&1.0&1.0&2.0&1,205.0\\
&EUSES&0&1.0&1.0&2.0&273.0\\
\midrule
c7 Total number of transitive precedents per formula&Enron&0&1.0&3.0&18.0&22,702.0\\
&EUSES&0&1.0&4.0&15.0&6,536.0\\
\bottomrule
\end{tabulary}
\end{minipage}
\end{table*}

\subsection{Use of Functions}

Modern spreadsheet software contains over 300 different functions that users can combine to create complex formulas. Furthermore, the functions form the building blocks for the spreadsheet model. Research on learning APIs has shown that analyzing usage of source code can reveal interesting patterns \cite{montandon2013documenting}. Therefore, we have analyzed the use of functions in the Enron dataset and compared it with EUSES.

From Hermans and  Murphy-Hill, we know that there is little diversity in the functions that are used in the Enron spreadsheets \cite{Hermans:2014aa}. In this analysis, we observe the same. We have selected all unique formulas with at least one function. For this metric we focussed on the use of build-in Excel functions (F1). Hence, we excluded operators (+, -, *, /, \^{}) and user-defined functions. In the resulting formulas the top 15 functions cover 69\% of the spreadsheets. For EUSES, this is almost the same: 70\%. 

Table \ref{top10ofmostfrequentlyusedfunctions} gives an overview of the fifteen most frequently used functions in the two datasets.

\begin{table}[htbp]
\begin{minipage}{\linewidth}
\setlength{\tymax}{\linewidth}
\centering
\small
\caption{Top 15 of most frequently used functionss}
\label{top10ofmostfrequentlyusedfunctions}
\begin{tabulary}{\textwidth}{lll}\toprule
&Enron&EUSES\\
\midrule
1&\textbf{SUM}&\textbf{SUM}\\
2&\textbf{IF}&\textbf{IF}\\
3&\textbf{AVERAGE}&\textbf{ROUND}\\
4&\textbf{VLOOKUP}&HYPERLINK\\
5&\textbf{ROUND}&\textbf{CONCATENATE}\\
6&SUBTOTAL&AND\\
7&OFFSET&COUNTIF\\
8&\textbf{CONCATENATE}&\textbf{AVERAGE}\\
9&NOW&OR\\
10&DAVERAGE&INDIRECT\\
11&SUMIF&MIN\\
12&INDEX&ISNUMBER\\
13&MATCH&MAX\\
14&LOOKUP&\textbf{VLOOKUP}\\
15&MONTH&ISBLANK\\
\bottomrule
\end{tabulary}
\end{minipage}
\end{table}

Although it is true that a small set of functions covers the majority of spreadsheets, this set of functions is not the same in each dataset. We see that only six functions (marked in bold) are present in both datasets. This probably indicates that the set of functions that are used in a spreadsheet depends on either the company, department, user or business domain.

In the previous paragraphs we have looked at the individual functions. However there are several functions that can be used for the same kind of problem. For example, VLOOKUP, HLOOKUP, MATCH, and SEARCH all have the purpose of looking up data. To better understand what users try to accomplish with functions, it would be useful to analyze the use of functions on the level of these groups (f2). To do so, we have used the classification that was defined by Microsoft\footnote{https://support.office.com/en-AU/Article/Excel-functions-by-category-5f91f4e9-7b42-46d2-9bd1-63f26a86c0eb}.

\begin{table}[htbp]
\begin{minipage}{\linewidth}
\setlength{\tymax}{\linewidth}
\centering
\small
\caption{Relative use of functions by category}
\label{relativeuseoffunctionsbycategory}
\begin{tabulary}{\textwidth}{lrr}\toprule
Category             &Enron  &EUSES\\
\midrule
Operator             &71.4\% &58.5\%\\
Math and trigonometry&16.4\% &19.2\%\\
Logical               &3.6\%  &9.9\%\\
Lookup and reference  &2.8\%  &3.6\%\\
Statistical           &2.5\%  &4.0\%\\
Date and time         &1.6\%  &0.5\%\\
Text                  &0.7\%  &1.8\%\\
Information           &0.5\%  &1.0\%\\
Database              &0.3\%  &0.1\%\\
Financial             &0.1\%  &0.4\%\\
Engineering           &0.0\%  &0.0\%\\
Compatibility         &0.0\%  &0.7\%\\
Udf                   &0.0\%  &0.4\%\\
User defined add-ins  &0.0\%  &0.0\%\\
\bottomrule
\end{tabulary}
\end{minipage}
\end{table}

 In Table \ref{relativeuseoffunctionsbycategory}, we see a list of these categories and the percentage of formulas that contains a function within this category. It is clear that operators are used the most frequently. If we shift our focus to the pure use of functions we see that the majority of functions belong to the category \emph{Math and trigonometry}. The functions SUM, ROUND and SUBTOTAL are responsible for 95\% of this category. We conclude that functions are used mostly to perform arithmetic calculations.
Another common category is \emph{Logical}. The most frequently used functions within this category are: IF, AND, and OR (used in 99,8\% of the formulas of the \emph{Logical} category). Users use these functions not to calculate, but to reason with formulas, often with the goal to check or prevent for errors.

\begin{table*}[htbp]
\begin{minipage}{\linewidth}
\setlength{\tymax}{\linewidth}
\centering
\small
\caption{Overview of formula characteristics}
\label{functionmetrics}
\begin{tabulary}{\textwidth}{llrrrrr}\toprule
Metric&Dataset&Min&Q1&Median&Q3&Max\\
\midrule
f3 Number of preceding cells per formula&Enron&0.0&1.0&2.0&4.0&13,057.0\\
&EUSES&0.0&1.0&2.0&4.0&3,698.0\\
\midrule
f4 Parse tree depth per formula&Enron&1.0&2.0&2.0&3.0&106.0\\
&EUSES&1.0&2.0&2.0&3.0&89.0\\
\midrule
f5 Number of unique functions per formula&Enron&1.0&1.0&1.0&1.0&8.0\\
&EUSES&1.0&1.0&1.0&1.0&8.0\\
\bottomrule
\end{tabulary}
\end{minipage}
\end{table*}

The category \emph{Lookup and reference} also catches the eye. It indicates the need of users to lookup or refer to data in their spreadsheets. The functions VLOOKUP, OFFSET, INDEX, LOOKUP, and MATCH, together, are responsible for 81\% of the formulas in this category.

Functions within the category \emph{Compatibility} are those that Microsoft has replaced with new functions, but because of compatibility reasons they are still available. Although the Enron dataset stems from 2000 and 2001, we observe that only a very small fraction (\textless 0.1\%) of formulas use a function that in the meantime has been replaced by Microsoft. This means that the majority of functions used in the Enron set are still valid and unchanged functions in the latest version of Excel.

Above we discussed the kind of functions that are used and the purpose of these functions. But what about complexity? How complex are formulas? A high number of references that are made within the formula to other cells (preceding cells) can make a formula difficult to understand. The same is true for the degree of nestedness of a formula, that we can measure with the parse tree depth of the formula. The results for both metrics can be found in Table \ref{functionmetrics}.

The number of preceding cells (f3) is almost identical for both Enron and EUSES. The differences are in the extremities. The high maxima for the number of precedents in both datasets are caused by references to a large range. For example the formula for the cell with the maximum number of preceding cells in the Enron set is:

\begin{verbatim}
=VLOOKUP(B51,[50]jan98!$A$34:$IV$84,3,0)
\end{verbatim}

It is a range of 256 columns (which was the maximum number of columns within a worksheet in Excel 2000) and 51 rows, which makes a total of 13,056 cells. The lookup value in B51 gives us the maximum of 13,057 preceding cells.
With a median of two preceding cells for all datasets, it is reasonable to assume that the number of preceding cells is not causing complexity.

In Table \ref{functionmetrics} we observe that only the maximum values for the parse tree depth (f4) differ between Enron and EUSES. One could also argue that a formula with a parse tree depth of two is not really complex. We were intrigued by the formula with the maximum parse tree depth of 106 in the Enron set. One would expect that such a formula is quite complex. However in this specific example the formula consisted of 105 hard coded numbers that were added with the + operand. So definitely a large formula, but not complex.

Another factor that could cause complexity is the number of different functions that is used in a formula (f5). In the majority of formulas, only one function is used. Formulas with more than three different functions are rare. Within the Enron set only 1.5\% of the formulas uses more than three different functions. For EUSES this percentages is even smaller (0.4\%).

To summarize our findings with respect to the use of functions, we can confirm the finding of Hermans and Murphy-Hill that users only use a very small set of functions\cite{Hermans:2014aa}. The composition of this set differs between the different dataset, indicating that the kind of functions are user-, company- or domain-dependent. Functions are mostly used to perform arithmetic calculations, add logic to spreadsheets, and lookup or reference data. We also observed that formulas are not that complex if you compare them to formulas found in \cite{hermans2014detecting}.

\section{Enron versus EUSES}

We used the Wilcoxon-Mann-Whitney test to determine if the characteristics of the Enron spreadsheet differ from the EUSES spreadsheets. The test calculates a p-value. The results can be found in Table \ref{enroncomparedtoeuses}.

\begin{table}[htbp]
\begin{minipage}{\linewidth}
\setlength{\tymax}{\linewidth}
\centering
\small
\caption{Enron compared to EUSES}
\label{enroncomparedtoeuses}
\begin{tabulary}{\textwidth}{lrr}\toprule
Metric&p-value&\emph{d}\\
\midrule
s1 number of cells&\textless 0.01&0.090\\
s2 number of worksheets&\textless 0.01&0.087\\
s3 number of formulas&\textless 0.01&0.136\\
s4 number of unique formulas&\textless 0.01&0.114\\
s5 length of formula&\textgreater 0.05&\\
\midrule
c2 number of external links&\textless 0.01&0.243\\
c4 number of intraworksheet connections&\textless 0.01&0.142\\
c6 path depth&\textgreater 0.05&\\
c7 number of transitive precedents&\textless 0.05&0.006\\
\midrule
f3 number of preceding cells&\textgreater 0.05\\
f4 parse tree depth&\textgreater 0.05\\
f5 number of unique functions&\textgreater 0.05&\\
\bottomrule
\end{tabulary}
\end{minipage}
\end{table}

The results of the Wilcoxon-Mann-Whitney test indicate that for some metrics (e.g. number of cells) there is a significant difference in the distributions of the Enron and EUSES datasets. For other metrics (e.g. path depth) they are equal. For all metrics with a significant difference the effect (calculated with the Cliff's Delta \emph{d}) is negligible (\emph{d} \textless{} 0.147), except for \emph{number of external links}, where the effect is small (0.147 \textless{} \emph{d} \textless{} 0.33). So, although there is a statistical significant difference in the distribution for some metrics between the Enron and EUSES dataset, the effect of this difference is negligible or small. Based on this we can conclude that the spreadsheets in the EUSES corpus are comparable with the Enron spreadsheets with respect to the metrics used in this paper. The full dataset and R scripts are available online\footnote{http://figshare.com/articles/Enron\_versus\_EUSES/1298994}. 

\section{Discussion}
In the previous section, we have 1) described the results of an analysis of the spreadsheets in the Enron dataset with respect to the dimensions size, coupling and use of functions and 2) compared the Enron spreadsheets with the EUSES spreadsheets. In this section, we discuss some issues that could affect the applicability and suitability of the approach used.

\subsection{User-defined functions}
Our analysis focused on the Microsof Excel built-in functions. We did not analyze the use of user-defined functions, because the use of it is rather limited. Only in 0.5\% of the Enron and 1.2\% of the EUSES spreadsheets that contain formulas with functions, user-defined functions were used.

\subsection{Pivot tables, charts and VBA code}
In this first detailed analysis of the Enron spreadsheets, we limited ourselves to metrics for size, coupling and use of functions. More elaborate constructs like Pivot tables, charts and VBA code could also have an impact on the complexity of spreadsheets. In future research, we plan to extend the current analysis with these constructs.

\subsection{Path analysis}
In this paper, we looked at the use of formula and functions on the level of the individual cell. However, formulas receive input from precedent cells. These precedent cells, in turn, can also contain formulas. One could argue that all the formulas in the calculation chain form a small program or method. In future work, we will analyze these programs to see if there are certain common patterns that can be found in the majority of spreadsheets.

\subsection{Threats to validity}

\subsubsection{Cleaned dataset}
Before the Enron dataset was made publicly available, it has been cleaned. In total over 10,000 e-mail messages with potential sensitive personal information, like credit card numbers, identity numbers, personal contact details, etc., were deleted from the dataset \cite{Cassidy:aa}. We do not believe that this cleaning affected the analysis of the spreadsheets. The spreadsheets themselves were most likely not deemed personal by the researchers cleaning the dataset.

\subsubsection{Age of dataset}
The spreadsheets in the Enron dataset are more than ten years old. However, we believe that the way users create spreadsheets has not change much over this period. This is supported by the fact that in the meanwhile the user interface of Microsoft Excel has not changed significantly.

\section{Related Work}

Most related to our efforts is of course the work of Hermans and Murphy-Hill \cite{Hermans:2014aa} which initially presented the Enron spreadsheet corpus. The authors performed a preliminary analysis of some basic characteristics (like number of worksheets, cells, and formulas) of these spreadsheets. While Hermans and Murphy-Hill present a first overview of the spreadsheets, we, in this paper, dive deeper and added additional metrics to measure the degree of coupling and gain more insight in the actual use of functions. Also, for every metric we compared the Enron spreadsheets with the EUSES corpus. By using the Wilcoxon-Mann-Whitney test and calculating the Cliff's delta for the different metrics we answer the question of representativeness of the EUSES corpus. 

Secondly, there is the EUSES corpus, which was introduced by Fisher and Rothermel in 2005 \cite{fisher2005euses}. Besides EUSES, there are a few other smaller corpora\cite{panko2000two,powell2008critical}. Unfortunately, none of these corpora were publicly available to include in the analyses of this paper.

In addition to work on spreadsheet corpora, there are papers on spreadsheet metrics, which have in common with our work that they too measure properties of spreadsheets\cite{breg2004, hodn2008}. There is an overlap between the metrics defined in these papers and the metrics we use in this analysis (e.g. number of cells, number of formulas). We have added metrics to obtain a better understanding of how functions are used within spreadsheets. We have also used the metrics to compare two spreadsheet corpora.   

\section{Concluding Remarks}

The overall goal of this paper is to understand and quantify the characteristics of industry grade spreadsheets and secondly to see if these spreadsheets differ from the EUSES corpus that was used frequently in previous research. To do so, we formulated three research questions:

\subsection{What are the characteristics, in terms of size and coupling, of spreadsheets in a business environment?}

\begin{itemize}
\item The results of the analysis indicate that the size of the majority of the Enron spreadsheets is small. On average they consists of 701 non-empty cells, contain three worksheets and have about 128 unique formulas.
\item For coupling, we see a similar picture. The majority of spreadsheets (89\%) is not linked to other spreadsheets and also within the spreadsheet only 20\% of them have links between worksheets within the spreadsheet. On the cell level, we see the same. The median path depth (calculation chain) is one and a cell with a formula has a median of three transitive precedent cells.
\item Based on the results of the analysis, we conclude that in a business environment large spreadsheets with a high degree of coupling do exist, but are not very common.
\end{itemize}

\subsection{How are spreadsheet functions used and combined?}

\begin{itemize}
\item With respect to the use of functions, we see that users use only a small subset of the available built-in functions. Surprising is that this subset of functions is not the same between the different datasets. This could be an indication that the kind of functions that are used in a spreadsheet are depending on the individual user or the specific business domain.
\item If we look at the level of function categories, we notice that mainly the general purpose categories like \emph{Operators}, \emph{Math and trigonometry}, \emph{Lookup and reference} and \emph{Logical} are used. Users tend to use functions to either calculate something, look up something or a combination of the two. Functions from the \emph{Logical} category are used to reason with formulas, probably with the goal to check or prevent for errors.
\item Regarding the complexity, the results show us that the majority of formulas are simple. They only make a direct reference to a few other cells (median: two), are hardly nested (median for the parse tree depth: two) and most of the times only contain one function.
\end{itemize}

\subsection{Is there a difference with respect to these metrics between spreadsheets that are used in a business context and the EUSES corpus?}

\begin{itemize}
\item With respect to size, coupling and use of functions the Enron spreadsheets are quite similar to the EUSES corpus. Although we find for most metrics a statistical significant difference, the effect size is either negligible (6 metrics) or small (1 metric). Based on this we can conclude that for these characteristics the EUSES corpus is representative for spreadsheets that are used in a business context.
\end{itemize} 

\section{Future work}

This paper gives rise to several directions for future work. The results of our analysis show that users in general create small spreadsheets, with a relatively low degree of coupling and that they use simple formulas. This makes it less likely that the complexity of a spreadsheet is causing the high error rate of spreadsheets. It still leaves us with the question, what is causing errors in spreadsheets? To answer this question, one of the directions for future research could be to focus on the user interface of the spreadsheet software. In previous work \cite{jansen2014using}, we introduced the concept of a visual language to model spreadsheets and in future research we are planning to develop and validate such a language.

In this paper we focused on formulas at the cell level. The formula in a spreadsheet could be considered as source code. However, by looking at a single cell you are actual looking at a single line of source code. In future research, we are planning to extract and combine all formulas in the same calculation chain to obtain a better and completer understanding of the use of functions within spreadsheets. This will allow us to search for commonly used patterns.

\bibliographystyle{IEEEtran}
\bibliography{IEEEabrv,/Users/basjansen/Dropbox/TUDelft/B/BibDesk/bib}

\end{document}